\documentclass[12pt]{article}
\usepackage[backend=biber,style=nature,sorting=none,block=ragged]{biblatex} 
\usepackage[T1]{fontenc}
\usepackage[none]{hyphenat}
\usepackage[title]{appendix}
\usepackage[affil-it]{authblk}
\usepackage{abstract}
\usepackage[svgnames,dvipsnames,x11names]{xcolor}
\usepackage{graphics,graphicx}
\usepackage[right=15mm,left=15mm,bottom=20mm,top=20mm]{geometry}
\usepackage{amsmath,amssymb,amsthm,amsfonts,upgreek
	,accents,cancel,bm}
\usepackage{enumitem}
\usepackage{physics}
\usepackage{euscript}
\usepackage{lmodern}
\usepackage[colorlinks=true,linkcolor=blue
,citecolor=red,urlcolor=purple]{hyperref}

\newcommand{\changelinkcolora}{\hypersetup
	{linkcolor={Aqua!60!black}}}

\newcommand{\changelinkcolorb}{\hypersetup
	{linkcolor={Green!70!black}}}

\newcommand{\changeurlcolor}{\hypersetup
	{urlcolor={Sepia}}}%

\DeclareMathOperator{\uppartialbold}
{{\raisebox{-1 pt}{\rotatebox{20}{$\boldsymbol\partial$}}}
	\hspace{-2 pt}}

\numberwithin{equation}{section}

\numberwithin{footnote}{section}

	\title{\textbf{Generators of Local Lorentz Transformation\\in ADM-Vielbein Formalism\\of Gravitational Relativity}}

	\author[a]{Alireza Faraji\footnote
		{\href{mailto:alireza.faraji@ph.iut.ac.ir}
			{alireza.faraji@ph.iut.ac.ir}}}
	
	\author[b]{Zahra Molaee\footnote
		{\href{mailto:zhrmolaee8@gmail.com}	
			{zhrmolaee8@gmail.com}}}
	
	\author[a]{Ahmad Shirzad\footnote
		{\textcolor{Sepia}{Corresponding author:} \href{mailto:shirzad@iut.ac.ir}
			{shirzad@iut.ac.ir}}}
	
	\affil[a]{{Department of Physics, Isfahan University of Technology, Iran}}
	
	\affil[b]{{School of Astronomy, Institute for Research in  Fundamental Sciences (IPM),\hspace{5 cm} P.O. Box  19395-5531, Tehran, Iran.}}
	
\begin{document}
		{\changeurlcolor
			\maketitle}
	
	\begin{abstract}
General relativity contains 16 variables in the framework of ADM-Vielbein formalism, which are six more than in the metric formalism. These variables emerge due to the additional symmetry of local Lorentz transformations. In the framework of the Hamiltonian approach, it is expected to find first class constraints that generate this gauge symmetry. We introduce the complete form of such constraints and show that they exactly obey the algebra of the Lorentz group.
\vspace{10 cm} 
	\end{abstract}
\pagebreak
\section{Introduction}
All covariant theories concerning the dynamics of spacetime may be formulated in terms of the \linebreak"metric field" as well as the "vielbein field"~\cite{carroll2019spacetime,padmanabhan2010gravitation}. The vielbein field contains 16 components in four dimensions, while the metric has 10 independent components. Six additional variables correspond to our freedom in choosing the local inertial frame. So, the vielbein formulation, regardless of which theory is taken into account, is characterized by the local gauge symmetry due to local Lorentz transformations (LLT), which are composed of rotation and boost symmetry in the tangent space.

On the other hand, the local symmetry due to diffeomorphism introduces four additional arbitrary gauge fields. In this way, every covariant theory in the vielbein formalism should at least support the ten-parameter gauge group of LLT~$\times$~diffeomorphism. Fortunately, these two groups are distinct and may be studied independently. As is well known, in the framework of Hamiltonian formalism, every local gauge symmetry, should be generated by a set of first class constraints~\cite{dirac2013lectures,henneaux1992quantization}. In fact, each gauge symmetry corresponds, in some sense, to one chain of first class constraints~\cite{loran2002classification}. 

The constraint structure of general relativity (GR) in the metric formalism has been well known for a long time~\cite{charap1983canonical,sundermeyer1982constrained}.
In fact, the momenta $\mathcal{P}_0$ and $\mathcal{P}_i$, conjugate to lapse and shift variables $N$ 
and $N^i$ respectively, are first level constraints. The consistency of $\mathcal{P}_0$ and $\mathcal{P}_i$ leads to second level constraints, known as Hamiltonian and momentum constraints, $\mathcal{C}$ and ${\mathcal C}_i$. It turns out that all these eight constraints are first class. Subtracting eight constraints and eight gauge fixing conditions from 20 phase space variables yields four phase space dynamical variables, which correspond to two degrees of freedom in configuration space.

In addition to the metric formalism, alternative formulations of general relativity exist that are based on different geometric structures, such as torsion. The approach adopted in this research follows the torsion-free formalism. A well-known alternative is the Teleparallel Equivalent of General Relativity (TEGR), in which the Lagrangian is constructed from the torsion tensor, resulting in an explicit breaking of local Lorentz invariance. For further details, see Ref.~\cite{aldrovandi2012teleparallel}. The structure of constraints and their algebra within this framework has been examined in several studies, including Refs.~\cite{maluf2001hamiltonian,maluf2013teleparallel}.

In the Hilbert-Einstein action, the spin connection is not an independent and is defined in terms of the tetrads. In contrast, the Palatini formulation of general relativity treats the spin connection and tetrads as independent variables. The Ashtekar-Barbero variables~\cite{ashtekar1986new,barbero1995real} are defined within this formalism, and the Holst action~\cite{holst1996barbero} is considered a generalization of the Palatini action. Refs.~\cite{montesinos2020canonical,montesinos2023linking,montesinos2018manifestly,montesinos2020canonicalHolst} include important developments such as the elimination of second class constraints and the construction of mappings between the ADM formalism and other formulations (and vice versa). However, the explicit structure of the Hamiltonian constraints and their algebra, which is the main focus of the present paper, has not been clearly addressed in those studies.
It is also important to emphasize that our approach is based on the Hilbert-Einstein action, in which the spin connection is not treated as an independent variable, unlike in the Palatini formulation.

Despite the various geometric structures used in general relativity, it remains important to investigate the existence of LLT symmetry within the ADM-Vielbein framework under torsion-free conditions. From the vielbein point of view, we should have six chains of first class constraints, responsible for generating the LLT gauge symmetry~\cite{heidari2020structure,batlle1986equivalence}. However, since we have only 6$\times$2 additional phase space variables, we cannot have more than 6 first class constraints due to LLT. This means that, in the Hamiltonian formalism, we should have only 6 first class constraints at the first level, which should generate this gauge symmetry. This is good news, which indicates that for LLT gauge symmetry, there is no issue regarding the consistency of the first level constraints. 

It is only necessary to identify 6 first class constraints out of the canonical momentum fields derived from the Lagrangian of the theory. However, the important point is that the algebra of the Poisson brackets of the constraints should be the same as the algebra of the generators of the Lorentz group. Hence, it is insufficient to show that the mutual Poisson brackets vanish weakly. Instead, one should show explicitly they obey the algebra of the Lorentz group, i.e., $\mathrm{SO(1,3)}$.

As far as we have searched, this simple task has not been clearly performed even for the simplest covariant theory of GR. In Ref.~\cite{charap1983canonical}, this procedure is carried out on the basis of the concept of proper time. However, explicit expressions concerning the relations of the vielbein and metric variables in the ADM formalism are not used, although the Lorentz algebra is fulfilled by the constraints. In fact, the explicit forms of vielbein variables in terms of lapse $N$, shift $N^i$, spatial vielbein ${e_i}^a$, and boost parameter $q_a$ (see the following {\changelinkcolora\hyperlink{section.2}{section}}) have only become known more recently since the work of Peldan~\cite{peldan1994actions} and others (see for instance~\cite{hinterbichler2012interacting}). This is precisely what we utilize in our Hamiltonian analysis in this paper.

A simplified version of the problem may be considered in which the boost parameters $q_a$ are set to zero~\cite{hinterbichler2012interacting}. In fact, this simplification fixes the boost symmetry and results in an upper triangular form of the vielbein. In this way, for a pure gravity theory, we are left with only thirteen variables $N$, $N^i$, and ${e_i}^a$, and the remaining symmetry is just a local rotation group in the tangent space. In this case, it is straightforward to introduce three generators of rotation symmetry by properly combining the primary constraints. This has been demonstrated in the literature~\cite{hinterbichler2012interacting}.  

In order to restore the full content of the theory, one can impose a Lorentz boost with parameters $q_a$ on the spatial indices of an upper triangular vielbein to obtain its most general form, involving 16 variables $N$, $N^i$, ${e_i}^a$, and $q_a$. One expects naturally that in this theory (with 16 variables) 6 first class constraints emerge which obey the Lorentz algebra. However, the surprising point is that in some famous references, such as~\cite{hinterbichler2012interacting,hassan2019interactions,klusovn2014hamiltonian,hinterbichler2015note,de2015vielbein}, it is claimed that even by using the full form of vielbein variables, the variables $q_a$ drop out from the Hilbert-Einstein action. This proposition implies that one is not able to identify appropriate constraints that generate boost transformations.

In this paper, we will show that this is not the case. As we will see, three boost variables $q_a$ are retained in the action; even more, their spatial and temporal derivatives also appear explicitly in the action. We will show that there exist suitable combinations of canonical momenta which (i) vanish identically when their definitions in terms of the velocities are used, and (ii) their Poisson brackets close according to Lorentz algebra. As we will see, it is not an easy task to express these first class constraints in terms of canonical variables. In fact, a detailed procedure must be followed to construct these constraints step by step. Although some simple parts of rotation generators are known in the literature, the full form of rotation generators has not yet been presented. More importantly, no idea about the form of boost generators has been given so far (see section {\changelinkcolora\ref{Generators of boosts}} below).

In the next {\changelinkcolora\hyperlink{section.2}{section}}, we will briefly  review basics of the vielbein formalism in terms of ADM variables (section {\changelinkcolora\ref{ADM-Vielbein formalism}} is the only reviewing part of the paper).
It should be noted that, in order to avoid heavy algebraic calculations, we will work in the so-called mini-superspace, where $N=1$ and $N^i=0$. This means that we aim to focus on LLT and disregard the diffeomorphism. We will write the Hilbert-Einstein action, as a typical example, in terms of 12 variables ${e_i}^a$ and $q_a$ along with their temporal and spatial derivatives. Then, in section {\changelinkcolora\ref{Hamiltonian analysis}}, we will introduce the explicit form of canonical momenta and try to find independent combinations among them (i.e., constraints) that vanish when expressed in terms of velocities.

In section {\changelinkcolora\ref{Rotation algebra}}, we try to find the generators of the Lorentz algebra among the constraints. We begin with the generators $L_{ab}$ of rotational symmetry and show that their algebra closes in the same way as the angular momentum algebra. As we will see, the explicit form of the $L_{ab}$'s, as well as the manner in which their algebra closes, is not obvious and requires considerable detail.

In order to complete the Lorentz algebra, we attempt, in Section {\changelinkcolora\ref{Generators of boosts}}, to construct the generators $L_{0a}$  as the boost generators. The explicit forms of the $L_{0a}$'s,  their algebra among themselves, and their algebra with the $L_{ab}$'s are complicated issues that will be explained subsequently.

\section{ADM-Vielbein formalism}\label{ADM-Vielbein formalism}
The ADM variables consist of the lapse function $N$, the shift functions $N^i$, and the spatial metric $h_{ij}$. In this formalism, the spacetime metric and its inverse can be expressed as~\cite{arnowitt1962dynamics,padmanabhan2010gravitation,castellani1982first}
 \begin{equation}
	g_{\mu\nu}=\begin{bmatrix}
		-N^2+N^lN_l & N_j\\
		N_i & h_{ij}
	\end{bmatrix},
	\label{equa2.1}
\end{equation}
\begin{equation}
	g^{\mu\nu}=\begin{bmatrix}
		-N^{-2} & N^{-2}N^j\\
		N^{-2}N^i & h^{ij}-N^{-2}N^iN^j
	\end{bmatrix}.
	\label{equa2.2}
\end{equation}
 Greek indices $\mu, \nu,\cdots $ are used for spacetime coordinates, while Latin indices $i, j, k,\cdots$ denote purely spatial coordinates.
 
 An orthogonal basis $\mathbf{E}_A$ may be chosen for the tangent space at an arbitrary point $P$ with coordinates $x^\mu$. The vielbein fields ${E_\mu}^A$ are defined as the components of the coordinate basis $\uppartialbold_\mu$ in terms of the orthogonal basis $\mathbf{E}_A$, i.e.,
 $\uppartialbold_\mu={E_\mu}^A\mathbf{E}_A$. Hence,
 $\uppartialbold_\mu \cdot\uppartialbold_\nu=g_{\mu\nu}$
 and $\mathbf{E}_A\cdot\mathbf{E}_B=\eta_{AB}$ imply
 \begin{equation}
 {E_\mu}^A{E_\nu}^B\eta_{AB}=g_{\mu\nu},
 \label{equa2.3}
 \end{equation} 
 where $\eta_{AB}=\mathrm{diag}(-1,1,1,1)$ is the flat spacetime metric. This relation indicates that the vielbein can be regarded as the square-root of the metric $g_{\mu\nu}$. A similar description applies to the cotangent space at point $P$. Assuming
 $\mathbf{dx}^\mu={E^\mu}_A\mathbf{E}^A$ where 
 $\mathbf{E}^A\cdot\mathbf{E}^B=\eta^{AB}$ and 
 $\mathbf{dx}^\mu\cdot\mathbf{dx}^\nu=g^{\mu\nu}$, we obtain 
 \begin{equation}
{E^\mu}_A{E^\nu}_B\eta^{AB}=g^{\mu\nu}.
\label{equa2.4}
 \end{equation}
 This shows that the inverse vielbein can be regarded as the square-root of the inverse metric $g^{\mu\nu}$. The fields ${E_\mu}^A$, ${E^\mu}_A$, 
  $E_{\mu A}\equiv g_{\mu\nu}{E^\nu}_A=\eta_{AB}{E_\mu}^B$ and
  $E^{\mu A}\equiv g^{\mu\nu}{E_\nu}^A=\eta^{AB}{E^\mu}_B$ can be used to raise or lower indices and to convert curved indices $\mu, \nu,\cdots $ to flat ones $A,B,\cdots$ and vice versa.
 
  Analogous definitions and relations apply to the three-dimensional spatial vielbeins ${e_i}^a$, ${e^i}_a$, $e_{ia}$, and $e^{ia}$, where $i,j,\cdots$ refer to curved spatial indices and $a,b,\cdots$ to flat ones. For example
   \begin{equation}
  	{e_i}^a{e_j}^b\eta_{ab}=h_{ij},
  	\hspace{3mm} \eta_{ab}=\text{diag}(1,1,1),
  	\label{equa2.5}
  \end{equation}
  \begin{equation}
  	{e^i}_a{e^j}_b\eta^{ab}=h^{ij},
  	\hspace{3mm} \eta^{ab}=\text{diag}(1,1,1).
  	\label{equa2.6}
  \end{equation}

One possible solution to Eqs.~\eqref{equa2.3} and \eqref{equa2.4}
 involving the lapse, shift, and spatial vielbeins, is given by~\cite{klusovn2014hamiltonian}
\begin{equation}
\widehat{E}_\mu\,\!^A=\begin{bmatrix}
N & N^i{e_i}^a\\	0 & {e_i}^a
\end{bmatrix},
\label{equa2.7}
\end{equation}
\begin{equation}
\widehat{E}^\mu\,\!_A=\begin{bmatrix}
{N}^{-1} & 0\\ -{N^i}{N}^{-1} & {e^i}_a
\end{bmatrix}.
\label{equa2.8}
\end{equation}
This form of vielbein is referred to as the upper and lower triangular vielbein, respectively, and it contains thirteen variables $N$, $N^i$, and ${e_i}^a$ or ${e^i}_a$. So far, local rotational symmetry concerning the flat spatial indices in the tangent (or cotangent) space has been preserved. As we know, the general vielbein field comprises sixteen components, while the upper triangular vielbein includes only thirteen nonzero ones.
If we restrict ourselves to the triangular form of the vielbein, only the rotation constraints emerge in the Hamiltonian analysis, while the boost constraints do not appear.  
 Due to local rotational symmetry, the ten metric components are functions of these thirteen components. However, this form does not exhibit full symmetry under Lorentz transformations.
 
 To construct a more general form of vielbein that includes three additional parameters corresponding to boost transformations in the tangent space, one may apply the following Lorentz boost transformations to the triangular vielbein
\begin{equation}
\Lambda{(q)^A}_B=\begin{bmatrix}
\gamma & q_b\\
q^a & \delta^a_b+{(1+\gamma)}^{-1} q^aq_b
\end{bmatrix},
\label{equa2.9}
\end{equation}
\begin{equation}
{\Lambda}^{-1}{(q)^A}_B=\begin{bmatrix}
\gamma & -q_b\\
-q^a & {\delta^a_b+(1+\gamma)^{-1}q^aq_b}
\end{bmatrix}.
\label{equb2.10}
\end{equation}
The $q$-variables are identified as the boost parameters, with
\begin{equation}
\gamma\equiv\sqrt{1+q^aq_a}.
\label{equb2.11}
\end{equation}
For example, the action of $\Lambda(\mathbf{q})$ on the column vector $(1,\mathbf{0})$ yields $(\gamma,\mathbf{q})$. It is important to note that boost transformations act on upper and lower flat indices as 
$N'^A=\Lambda{(q)^A}_BN^B$ and 
$N'_A={\Lambda}^{-1}{(q)^B}_AN_B$, respectively.
Therefore, the generalized vielbein fields are given by
\begin{equation}
{{E}_\mu}^A=\Lambda{(q)^A}_B\widehat{E}_\mu\,\!^B,
\qquad{{E}^\mu}_A={\Lambda}^{-1}{(q)^B}_A\widehat{E}^\mu\,\!_B,
\label{equb2.12}
\end{equation}
which yield
\begin{equation}
{{E}_\mu}^A=\begin{bmatrix}
N\gamma+N^j{e_j}^cq_c&\,
Nq^a+N^j{e_j}^c\left(\delta_c^a
+{(1+\gamma)}^{-1}q_cq^a\right)\\
{e_i}^cq_c &\,  {e_i}^c\left(\delta_c^a+{(1+\gamma)}^{-1}q_cq^a\right)
\end{bmatrix},
\label{equb2.13}
\end{equation}
\begin{equation}
{{E}^\mu}_A=\begin{bmatrix}
N^{-1}\gamma &\, -N^{-1}q_a\\
-\left(N^iN^{-1}\gamma +q^c{e^i}_c\right) &\,  
N^iN^{-1}q_a+ {e^i}_c\left(\delta_a^c+{(1+\gamma)}^{-1}q_aq^c\right)
\end{bmatrix}.
\label{equb2.14}
\end{equation}
It is straightforward to verify that Eqs.~\eqref{equa2.3} and \eqref{equa2.4}
follow directly from Eqs.~\eqref{equb2.12}. These generalized tetrads, while still satisfying Eqs.~\eqref{equa2.3} and \eqref{equa2.4}, introduce three additional dynamical variables, thereby enabling the derivation of the boost constraints. It is worth noting that the tetrads in Eqs.~\eqref{equb2.13} and \eqref{equb2.14} represent the most general form, incorporating all six parameters associated with LLT, and they satisfy conditions \eqref{equa2.3} and \eqref{equa2.4}.

Our next task in this preliminary section is to recall the Hilbert-Einstein action in terms of ADM-Vielbein variables. To this end, let us define the following tensor constructed from vielbein variables
 \begin{equation}
	\Omega^{ABC}=E^{\mu A}{E^{\nu_B}}{\partial_{[\mu}
		{E_{\nu]}}^C}.
	\label{equb2.15}
\end{equation}
The spin connection can be expressed in terms of this tensor as follows
\begin{equation}
	{\omega_{\alpha}}^{AB}=\frac{1}{2}E_{\alpha C}
	(\Omega^{CAB}+\Omega^{BCA}-\Omega^{ABC}).
	\label{equb2.16}
\end{equation}
Following straightforward calculations (see Ref.~\cite{peldan1994actions}), it can be shown that the Hilbert-Einstein action, 
neglecting boundary terms, takes the following form
\begin{align}
S_{EH}&=\int\mathrm{d}^4xER\notag\\
&=\int\mathrm{d}^4xNe\left(\frac{1}{4}\Omega^{ABC}\Omega_{ABC}
+\frac{1}{2}\Omega^{ABC}\Omega_{ACB}-
{\Omega_{AC}}^A{\Omega_B}^{CB}\right),
\label{equb2.17}
\end{align}
where $E=Ne$ is the determinant of ${E_\mu}^A$ and $e$ is the determinant of ${e_i}^a$. As it is evident from Eq.~\eqref{equb2.17}, the Lagrangian $\mathcal{L}_{EH}$ is a function of $N$, $N^i$, ${e_i}^a$, $q_a$, and their spacetime derivatives. The Lagrangian in Eq.~\eqref{equb2.17}, although equivalent to the Hilbert-Einstein Lagrangian up to a surface term, is intrinsically invariant under local Lorentz transformations. This invariance arises from the tetrad structure of the theory.
 Since our main interest in this paper is to study LLT, we may simplify the problem in the other sector, i.e., diffeomorphism. To do this, we reduce the problem to the so-called mini-superspace, in which $N=1$ and $N^i=0$. Fortunately, this simplification considerably reduces the size of expressions and highlights the features related to LLT. With this simplification, $N=1$ implies $E=e$. The metric fields are given by
 \begin{equation}
g_{\mu\nu}=\begin{bmatrix}
-1 & 0\\
0 & h_{ij}
\end{bmatrix},
\hspace{10mm}
g^{\mu\nu}=\begin{bmatrix}
-1 & 0\\
0 & h^{ij}
\end{bmatrix}.
\label{equb2.18}
\end{equation}
The vielbein fields take the form
\begin{equation}
{{E}_\mu}^A=\begin{bmatrix}
\gamma&
q^a\\
{e_i}^cq_c & {e_i}^a+{(1+\gamma)}^{-1}{e_i}^cq_cq^a
\end{bmatrix},
\hspace{10mm}
{{E}^\mu}_A=\begin{bmatrix}
\gamma & -q_a\\
-q^c{e^i}_c &  
{e^i}_a+{(1+\gamma)}^{-1}{e^i}_cq^cq_a
\end{bmatrix}.
\label{equb2.19}
\end{equation}
The explicit forms of $\Omega^{ABC}$ and $\mathcal{L}_{EH}$ are involved and are given in 
{\changelinkcolorb\ref{Appendix omegaABC}} and {\changelinkcolorb\ref{Appendix l_EH}}.
\section{ Hamiltonian analysis}\label{Hamiltonian analysis}
\subsection{Momenta and constraints}
Upon fixing the variables $N$ and $N^i$, the Lagrangian $\mathcal{L}_{EH}$ becomes a function of ${e_i}^a$ and $q_a$. The first step is to determine the canonical momenta
${\pi^i}_a\equiv
\dfrac{\partial\mathcal{L}_{EH}}
{\partial\left(\partial_0{e_i}^a\right)}$ and 
$k^a\equiv
\dfrac{\partial\mathcal{L}_{EH}}
{\partial(\partial_0q_a)}$ 
conjugate to ${e_i}^a$ and $q_a$ respectively. Direct calculations, using the data given in {\changelinkcolorb\ref{Appendix pi^i_a}} and {\changelinkcolorb\ref{Appendix k^a}}, yield
\begin{align}
	{\pi^i}_a&=
	e\left(\partial_0{e^i}_a-e^{ib}{e^j}_b\partial_0e_{ja}
	+2{e^i}_a{e^j}_b\partial_0 {e_j}^b
	+2{e^j}_a{e^i}_b\partial_jq^b-2{e^i}_a{e^j}_b\partial_jq^b\right.
	\notag\\&\hspace{9 mm}
	\left. -2\gamma^{-1}(\gamma+1)^{-1}
	{e^j}_aq^c{e^i}_cq_b\partial_jq^b
	+2\gamma^{-1}(\gamma+1)^{-1}
	{e^i}_aq^b{e^j}_bq_c\partial_jq^c \right)	,
	\label{equc3.1}
\end{align}
and
\begin{align}
	k^a&=e\left(2\gamma^{-1}(\gamma+1)^{-1}e^{ja}q_b\partial_jq^b
	+2\partial_je^{ja}
	-2\gamma^{-1}(\gamma+1)^{-1}q^a{e^j}_b\partial_jq^b
	+2e^{ja}{e^k}_b\partial_j{e_k}^b \right.
	\notag\\&\hspace{9 mm}
	\left.-2\gamma^{-1}(\gamma+1)^{-1}q^aq^b\partial_j{e^j}_b
	-2\gamma^{-1}(\gamma+1)^{-1}q^aq^b{e^j}_b{e^k}_c\partial_j{e_k}^c\right).
	\label{equc3.2}
\end{align}
In order to find primary constraints, we need to identify identities among momenta (as well as the main variables) that are independent of the velocities. Fortunately, the momenta $k^a$ in Eq.~\eqref{equc3.2} do not involve velocities. Hence, three constraints emerge by transferring all terms on the right-hand side of Eq.~\eqref{equc3.2} to the left-hand side. However, this is not the only possible choice. We can construct any function we wish from the momenta $k^a$, and then impose the identities \eqref{equc3.2} to construct suitable first class constraints, which may be used to generate the Lorentz algebra of LLT gauge symmetry. Before proceeding further, it is important to emphasize once more that (despite the claim of some famous references) the variables $q_a$, as well as their canonical momenta, do appear in the theory, as they should.

As mentioned earlier, we need to find six first class constraints by using Eqs.~\eqref{equc3.1} and \eqref{equc3.2}. Let us begin with the former and try to construct constraints that generate rotations on the canonical variables ${e_i}^a$ and ${\pi^i}_a$. It is known that the functions $\tilde{M}_{ab}=e_{ia}{\pi^i}_b-e_{ib}{\pi^i}_a$ behave as angular momentum generators when acting on the vector indices "$a$" of the phase space variables ${e_i}^a$ and ${\pi^i}_a$. Using Eq.~\eqref{equc3.1}, the expression takes the following form
\begin{align}
	e_{ia}{\pi^i}_b-e_{ib}{\pi^i}_a&=
	e\left(e_{ia}\partial_0{e^i}_b-{e^i}_a\partial_0e_{ib}
	+2{e^i}_b\partial_iq_a
	-e_{ib}\partial_0{e^i}_a+{e^i}_b\partial_0e_{ia}
	-2{e^i}_a\partial_iq_b \right.
	\notag\\&\hspace{9 mm}
	\left.-2\gamma^{-1}(\gamma+1)^{-1}{e^i}_bq_aq_c\partial_iq^c
	+2\gamma^{-1}(\gamma+1)^{-1}{e^i}_aq_bq_c\partial_iq^c\right).
	\label{equc3.3}
\end{align}
Noticing the following identities
\begin{gather}
	e_{ia}\partial_0{e^i}_b+{e^i}_b\partial_0e_{ia}
	=\partial_0(e_{ia}{e^i}_b)=\partial_0\eta_{ab}=0,\notag\\
	-e_{ib}\partial_0{e^i}_a-{e^i}_a\partial_0e_{ib}
	=-\partial_0(e_{ib}{e^i}_a)=-\partial_0\eta_{ab}=0,
	\label{equc3.4}
\end{gather}
we have 
\begin{equation}
	{M}_{ab}\equiv\tilde{M}_{ab}+2eu_{ab}\approx0,
	\label{equc3.5}
\end{equation}
where
\begin{gather}
	u_{ab}\equiv{e^i}_a\partial_iq_b-{e^i}_b\partial_iq_a
	+\gamma^{-1}(\gamma+1)^{-1}q_c\partial_iq^c\left({e^i}_bq_a
	-{e^i}_aq_b\right).
	\label{equc3.6}
\end{gather}
\subsection{Notation and basic algebra}
At this stage, we are prepared to carry out lengthy calculations involving Poisson brackets. In order to shorten long expressions and improve clarity, we introduce the following abbreviations
\begin{gather}
f(\mathbf{x},t)=f,\qquad f(\mathbf{x'},t)=f',
\label{equc3.7}\\
\frac{\partial }{\partial x'^j}=\partial'_j,
\label{equc3.8}\\
\delta(\mathbf{x}-\mathbf{x'})=\delta_{\mathbf{x}\mathbf{x'}}.
\label{equc3.9}
\end{gather}
With these notations, the following identities hold
\begin{gather}
\partial'_j\delta_{\mathbf{x}\mathbf{x'}}=
-\partial_j\delta_{\mathbf{x}\mathbf{x'}},
\label{equd3.10}\\
f'\delta_{\mathbf{x}\mathbf{x'}}=
f\delta_{\mathbf{x}\mathbf{x'}},
\label{equd3.11}\\
fg'\partial_j\delta_{\mathbf{x}\mathbf{x'}}
=f\partial_j(g\delta_{\mathbf{x}\mathbf{x'}})
=fg\partial_j\delta_{\mathbf{x}\mathbf{x'}}
+f\partial_jg\delta_{\mathbf{x}\mathbf{x'}},
\label{equd3.12}\\
fg'\partial'_j\delta_{\mathbf{x}\mathbf{x'}}
=-fg'\partial_j\delta_{\mathbf{x}\mathbf{x'}}
=-fg\partial_j\delta_{\mathbf{x}\mathbf{x'}}
-f\partial_jg\delta_{\mathbf{x}\mathbf{x'}}.
\label{equd3.13}
\end{gather}
In this way, the fundamental Poisson brackets read
\begin{gather}
\left\{{e_i}^a,{{\pi'^j}_b}\right\}
=\delta_i^j\delta_b^a\delta_{\mathbf{x}\mathbf{x'}},\label{equd3.14}\\
\left\{q_a,k'^b\right\}=\delta_a^b\delta_{\mathbf{x}\mathbf{x'}}.
\label{equd3.15}
\end{gather}
A variety of Poisson brackets involving different vielbein variables, their conjugate momenta, and their spatial derivatives arise in our calculations. The following identities are particularly useful in subsequent manipulations
\begin{align}
&\left\{e_{ia},{\pi'^{jb}}\right\}
=\delta_i^j\delta_a^b\delta_{\mathbf{x}\mathbf{x'}},\label{equd3.16}\\
&\left\{e_{ia},{{\pi'^j}_b}\right\}
=\eta_{ab}\delta_i^j\delta_{\mathbf{x}\mathbf{x'}},\label{equd3.17}\\
&\left\{{e_i}^a,{\pi'^{jb}}\right\}
=\eta^{ab}\delta_i^j\delta_{\mathbf{x}\mathbf{x'}},\label{equd3.18}\\
&\left\{{e^i}_a,{\pi'^j}_b\right\}
=-{e^i}_b{e^j}_a\delta_{\mathbf{x}\mathbf{x'}},\label{equd3.19}\\
&\left\{e,{\pi'^i}_a\right\}
=e{e^i}_a\delta_{\mathbf{x}\mathbf{x'}},\label{equd3.20}\\
&\left\{\partial_k{e_i}^a,{{\pi'^j}_b}\right\}
=\delta_i^j\delta_b^a\partial_k\delta_{\mathbf{x}\mathbf{x'}}
,\label{equd3.21}\\
&\left\{\partial'_k{e'_i}^a,{{\pi^j}_b}\right\}
=-\delta_i^j\delta_b^a\partial_k\delta_{\mathbf{x}\mathbf{x'}}
,\label{equd3.22}\\
&\left\{\partial_k{e^i}_a,{\pi'^j}_b\right\}=
-\partial_k\left({e^i}_b{e^j}_a\right)
\delta_{\mathbf{x}\mathbf{x'}}
-{e^i}_b{e^j}_a\partial_k\delta_{\mathbf{x}\mathbf{x'}},\label{equd3.23}\\
&\left\{\partial'_k{e'^i}_a,{\pi^j}_b\right\}=
{e^i}_b{e^j}_a\partial_k\delta_{\mathbf{x}\mathbf{x'}},\label{equd3.24}\\
&\left\{q_a,k'_b\right\}=\eta_{ab}\delta_{\mathbf{x}\mathbf{x'}}
,\label{equd3.25}\\
&\left\{\partial_iq_a,k'_b\right\}
=\eta_{ab}\partial_i\delta_{\mathbf{x}\mathbf{x'}},\label{equd3.26}\\
&\left\{\partial'_iq'_a,k_b\right\}
=-\eta_{ab}\partial_i\delta_{\mathbf{x}\mathbf{x'}}.\label{equd3.27}
\end{align}
Most of the above identities follow straightforwardly. However, some of them involve subtleties and are derived in detail in {\changelinkcolorb\ref{Appendix two pb}}.
\section{Rotation algebra}\label{Rotation algebra}
\subsection{Algebra of \texorpdfstring{$\mathbf{M_{ab}}$}{PDFstring}'s}\label{Mab}
As we observed, the real constraints $M_{ab}$ are distinct from $\tilde{M}_{ab}$ which are preliminary generators of rotations. It is a straightforward exercise to verify that the tensor functions $\tilde{M}_{ab}$ satisfy the Lorentz algebra
\begin{align}
	\left\{\tilde{M}_{ab},\tilde{M}'_{cd}\right\}&=
	\left(\eta_{ac}\tilde{M}_{bd}-\eta_{bc}\tilde{M}_{ad}
	+\eta_{bd}\tilde{M}_{ac}-\eta_{ad}\tilde{M}_{bc}\right)
	\delta_{\mathbf{x}\mathbf{x'}}.
	\label{eque4.1}
\end{align}
From Eq.~\eqref{equc3.5} it follows that
\begin{align}
	\left\{{M}_{ab},{M}'_{cd}\right\}&=
	\left\{\tilde{M}_{ab},\tilde{M}'_{cd}\right\}
	+2e'\left\{\tilde{M}_{ab},u'_{cd}\right\}
	+2u'_{cd}\left\{\tilde{M}_{ab},e'\right\}
	\notag\\&\quad
	-2e\left\{\tilde{M}'_{cd},u_{ab}\right\}
	-2u_{ab}\left\{\tilde{M}'_{cd},e\right\}
	+4ee'\underbrace{\left\{u_{ab},u'_{cd}\right\}}_0.
	\label{eque4.2}
\end{align}
First, observe that
\begin{align}
	\left\{\tilde{M}_{ab},e'\right\}&=
	\left\{e_{ia}{\pi^i}_b-e_{ib}{\pi^i}_a,e'\right\}
	\notag\\&=
	-ee_{ia}{e^i}_b\delta_{\mathbf{x}\mathbf{x'}}
	+ee_{ib}{e^i}_a\delta_{\mathbf{x}\mathbf{x'}}
	\notag\\
	&=e(-\eta_{ab}+\eta_{ab})\delta_{\mathbf{x}\mathbf{x'}}
	\notag\\
	&=0,
	\label{eque4.3}
\end{align} 
and similarly,
\begin{equation}
\left\{\tilde{M}'_{cd},e\right\}=0.
\label{eque4.4}
\end{equation}
Fortunately, $u_{ab}$'s are coordinate-dependent functions only, so their mutual Poisson bracket vanishes. The only remaining task is calculating the brackets of the form $\pb{\tilde{M}}{u}$
\begin{align}
	2e'\left\{\tilde{M}_{ab},u'_{cd}\right\}&=
	2e'\left\{e_{ia}{\pi^i}_b-e_{ib}{\pi^i}_a
	,{{e'^j}_c\partial'_jq'_d-e'^j}_d\partial'_jq'_c
	+\gamma'^{-1}(\gamma'+1)^{-1}q'_e\partial'_jq'^e
	\left({e'^j}_dq'_c-{e'^j}_cq'_d\right)\right\}
	\notag\\&=
	2e\left(\gamma^{-1}(\gamma+1)^{-1}q_e\partial_iq^e
	\left(\eta_{ad}{e^i}_bq_c
	-\eta_{ac}{e^i}_bq_d
	-\eta_{bd}{e^i}_aq_c
	+\eta_{bc}{e^i}_aq_d\right) \right.
	\notag\\&\hspace{12 mm}
	\left. +\eta_{ac}{e^i}_b\partial_iq_d
	-\eta_{ad}{e^i}_b\partial_iq_c
	-\eta_{bc}{e^i}_a\partial_iq_d
	+\eta_{bd}{e^i}_a\partial_iq_c
	\right)
	\delta_{\mathbf{x}\mathbf{x'}}.
	\label{eque4.5}
\end{align}
Similarly, by interchanging $a\leftrightarrow c$ and $b\leftrightarrow d$, we have
\begin{align}	
	-2e\left\{\tilde{M}'_{cd},u_{ab}\right\}
	&=2e\left(\gamma^{-1}(\gamma+1)^{-1}q_e\partial_iq^e
	\left(-\eta_{bc}{e^i}_dq_a
	+\eta_{ac}{e^i}_dq_b
	+\eta_{bd}{e^i}_cq_a
	-\eta_{ad}{e^i}_cq_b\right) \right.
	\notag\\&\hspace{12 mm}
	\left. -\eta_{ac}{e^i}_d\partial_iq_b
	+\eta_{bc}{e^i}_d\partial_iq_a
	+\eta_{ad}{e^i}_c\partial_iq_b
	-\eta_{bd}{e^i}_c\partial_iq_a
	\right)\delta_{\mathbf{x}\mathbf{x'}}.
	\label{eque4.6}
\end{align}
Hence, we find
\begin{align}
	2e'\left\{\tilde{M}_{ab},u'_{cd}\right\}-2e\left\{\tilde{M}'_{cd},u_{ab}\right\}&=
	2e\left(\eta_{ac}u_{bd}-\eta_{bc}u_{ad}
	+\eta_{bd}u_{ac}-\eta_{ad}u_{bc}\right)
	\delta_{\mathbf{x}\mathbf{x'}}.
	\label{eque4.7}
\end{align}
Combining all results from Eqs.~\eqref{eque4.3}, \eqref{eque4.4}, \eqref{eque4.1}, and 
\eqref{eque4.7}, we finally arrive at
\begin{align}
	\left\{{M}_{ab},{M}'_{cd}\right\}=
	\left(\eta_{ac}{M}_{bd}-\eta_{bc}{M}_{ad}
	+\eta_{bd}{M}_{ac}-\eta_{ad}{M}_{bc}\right)
	\delta_{\mathbf{x}\mathbf{x'}}.
	\label{eque4.8}
\end{align}
This again reproduces the rotation sector of the Lorentz algebra.

However, despite satisfying the Lorentz algebra, the constraints  $M_{ab}$ are not yet the complete generators of the rotation group, since we must also generate rotations on the variables $q_a$ and $k^a$.
\subsection{Complete generators of rotations}\label{Lab}
In order to generate spatial rotations on the variables $q_a$ and $k^a$ we require generators of the form
\begin{equation}‌
\tilde{N}_{ab}=q_ak_b-q_bk_a,
\label{eque4.9}
\end{equation}
which clearly satisfy the Lorentz algebra
\begin{equation}
\left\{\tilde{N}_{ab},\tilde{N'}_{cd}\right\}=
\left(\eta_{ac}\tilde{N}_{bd}-\eta_{bc}\tilde{N}_{ad}
+\eta_{bd}\tilde{N}_{ac}-\eta_{ad}\tilde{N}_{bc}\right)
\delta_{\mathbf{x}\mathbf{x'}}.
\label{equf4.10}
\end{equation}
However, there is a subtle issue $\tilde{N}_{ab}$'s are not constraints. Using the definition \eqref{equc3.2} of momenta $k^a$, we find
\begin{align}
\tilde{N}_{ab}&=
-2e\left(\gamma^{-1}(\gamma+1)^{-1}
\left(q_b{e^i}_a-q_a{e^i}_b\right)q_c\partial_iq^c
+q_b\partial_i{e^i}_a-q_a\partial_i{e^i}_b
+\left(q_b{e^i}_a-q_a{e^i}_b\right)
{e^j}_c\partial_i{e_j}^c\right).
\label{equf4.11}	
\end{align}
Therefore, we must define the following modified generators
\begin{equation}
N_{ab}\equiv\tilde{N}_{ab}+2ew_{ab}\approx0,
\label{equf4.12}
\end{equation}
where
\begin{equation}
w_{ab}=\gamma^{-1}(\gamma+1)^{-1}
\left(q_b{e^i}_a-q_a{e^i}_b\right)q_c\partial_iq^c
+q_b\partial_i{e^i}_a-q_a\partial_i{e^i}_b
+\left(q_b{e^i}_a-q_a{e^i}_b\right)
{e^j}_c\partial_i{e_j}^c.
\label{equf4.13}
\end{equation}
Now, $N_{ab}$ are actual constraints that are expected to satisfy the Lorentz algebra. However, to save time, it is convenient to work with the total generators defined as
\begin{equation}
L_{ab}=M_{ab}+N_{ab}.
\label{equf4.14}
\end{equation}
Assuming
\begin{gather}
	L_{ab}=\tilde{L}_{ab}+2v_{ab},\label{equf4.15}\\
	 \tilde L_{ab}=\tilde M_{ab}+\tilde N_{ab},
	\label{equf4.16}
\end{gather}
and using $\partial_ie=e{e^j}_c\partial_i{e_j}^c$,
we obtain
\begin{align}
	v_{ab}=\partial_i\left(e\left(q_b{e^i}_a-q_a{e^i}_b\right)\right).
	\label{equf4.17}
\end{align}
It is now evident that the generators $L_{ab}$ are the appropriate constraints that generate spatial rotations on the variables ${e_i}^a$ and $q^a$. The next step is to verify explicitly that they indeed satisfy the Lorentz algebra. 
\subsection{Rotation algebra of \texorpdfstring{$\mathbf{L_{ab}}$}{PDFstring}'s}\label{Algebra Lab}
In this subsection, we aim to investigate the algebra of the generators ${L_{ab}}$. According to Eqs.~\eqref{equf4.15} and \eqref{equf4.16}, we obtain
\begin{align}
\left\{{L}_{ab},{L}'_{cd}\right\}&=
\left\{\tilde{L}_{ab},\tilde{L}'_{cd}\right\}
+2\left\{\tilde{L}_{ab},v'_{cd}\right\}
-2\left\{\tilde{L}'_{cd},v_{ab}\right\}
+4\left\{v_{ab},v'_{cd}\right\}.
\label{equf4.18}
\end{align}
Using Eqs.~\eqref{equf4.16}, \eqref{eque4.1}, and \eqref{equf4.10} it is evident that the ${\tilde L_{ab}}$'s satisfy the Lorentz algebra. Since $\left\{v_{ab},v'_{cd}\right\}~=0$, it suffices to calculate
\begin{equation}
\left\{\tilde{L}_{ab},v'_{cd}\right\}=
\left\{\tilde{M}_{ab},v'_{cd}\right\}
+\left\{\tilde{N}_{ab},v'_{cd}\right\}.
\label{equf4.19}
\end{equation}
We calculate the first term straightforwardly
\begin{align}
\left\{\tilde{M}_{ab},v'_{cd}\right\}&=
\left\{\tilde{M}_{ab},\partial'_j
\left(e'\left(q'_d{e'^j}_c-q'_c{e'^j}_d\right)\right)\right\}
\notag\\&=
\partial'_j\left(e'\left\{\tilde{M}_{ab},q'_d{e'^j}_c-q'_c{e'^j}_d\right\}\right)
\notag\\&=
\partial'_i\left(e\left(\eta_{ac}q_d{e^i}_b
-\eta_{ad}q_c{e^i}_b-\eta_{bc}q_d{e^i}_a
+\eta_{bd}q_c{e^i}_a\right)
\delta_{\mathbf{x}\mathbf{x'}}\right)
\label{equf4.20}.
\end{align}
For the second term, we find
\begin{align}
\left\{\tilde{N}_{ab},v'_{cd}\right\}&=
\partial'_i\left(e'\left\{q_ak_b-q_bk_a,
q'_d{e'^i}_c-q'_c{e'^i}_d \right\}\right)
\notag\\&=
\partial'_i\left(e\left(-\eta_{bd}q_a{e^i}_c
+\eta_{bc}q_a{e^i}_d
+\eta_{ad}q_b{e^i}_c-\eta_{ac}q_b{e^i}_d\right)
\delta_{\mathbf{x}\mathbf{x'}}\right).
\label{equf4.21}
\end{align}
Here, we have used Eq.~\eqref{eque4.3} and the analogous identity for $\tilde{N}_{ab}$. By adding Eqs.~\eqref{equf4.20} and \eqref{equf4.21}, we obtain
\begin{align}
\left\{\tilde{L}_{ab},v'_{cd}\right\}&=
-e\left(\eta_{bd}\left(q_c{e^i}_a-q_a{e^i}_c\right)
-\eta_{bc}\left(q_d{e^i}_a-q_a{e^i}_d\right) \right.
\notag\\&\hspace{13 mm}
\left. +\eta_{ac}\left(q_d{e^i}_b-q_b{e^i}_d\right)
-\eta_{ad}\left(q_c{e^i}_b-q_b{e^i}_c\right)\right)\partial_i
\delta_{\mathbf{x}\mathbf{x'}}.
\label{equf4.22}
\end{align}
By interchanging $a\leftrightarrow c$, $b\leftrightarrow d$,
and replacing $\partial'\rightarrow \partial$, we similarly find
\begin{align}
-\left\{\tilde{L}'_{cd},v_{ab}\right\}&=
-\partial_i\left(e\left(\eta_{bd}
\left(q_a{e^i}_c-q_c{e^i}_a\right)
-\eta_{ad}\left(q_b{e^i}_c-q_c{e^i}_b\right) \right.\right.
\notag\\&\hspace{14 mm}
\left.\left.+\eta_{ac}\left(q_b{e^i}_d-q_d{e^i}_b\right)
-\eta_{bc}\left(q_a{e^i}_d-q_d{e^i}_a\right)\right)
\delta_{\mathbf{x}\mathbf{x'}}\right).
\label{equf4.23}
\end{align}
Using Eqs.~\eqref{equf4.22}, \eqref{equf4.23}, and the definition of $v_{ab}$ in Eq.~\eqref{equf4.17}, we conclude
\begin{equation}
2\left\{\tilde{L}_{ab},v'_{cd}\right\}
-2\left\{\tilde{L}'_{cd},v_{ab}\right\}=
2\left(\eta_{ac}v_{bd}-\eta_{bc}v_{ad}
+\eta_{bd}v_{ac}-\eta_{ad}v_{bc}\right)
\delta_{\mathbf{x}\mathbf{x'}}.
\label{equf4.24}
\end{equation}
Finally, combining Eqs.~\eqref{equf4.18} and \eqref{equf4.24}, we find that the generators $L_{ab}$ indeed satisfy the rotation part of the Lorentz algebra
\begin{equation}
\{{L}_{ab},{L}'_{cd}\}=
\left(\eta_{ac}{L}_{bd}-\eta_{bc}{L}_{ad}
+\eta_{bd}{L}_{ac}-\eta_{ad}{L}_{bc}\right)
\delta_{\mathbf{x}\mathbf{x'}}.
\label{equf4.25}
\end{equation}
\section{Generators of boosts}\label{Generators of boosts}
\subsection{Preliminary construction}\label{tilde L0a}
At this point, we aim to extend the rotation algebra \eqref{equf4.25} to the full Lorentz algebra, given by
\begin{equation}
	\{{L}_{AB},{L}'_{CD}\}=
	\left(\eta_{AC}{L}_{BD}-\eta_{BC}{L}_{AD}
	+\eta_{BD}{L}_{AC}-\eta_{AD}{L}_{BC}\right)
	\delta_{\mathbf{x}\mathbf{x'}}.
	\label{equg5.1}
\end{equation}
To achieve this, we need to construct three weakly vanishing quantities $L_{0a}$ such that 
  \begin{gather}
	\left\{{L}_{0a},{L}'_{0b}\right\}=
	-{L}_{ab}
	\delta_{\mathbf{x}\mathbf{x'}},
	\label{equg5.2}\\
	\left\{{L}_{0a},{L}'_{cd}\right\}=
	\left(\eta_{ad}{L}_{0c}-\eta_{ac}{L}_{0d}\right)
	\delta_{\mathbf{x}\mathbf{x'}}.
	\label{equg5.3}
\end{gather}
To this end, we first define "preliminary boost generator" $\tilde L_{0a}$ as appropriate combinations of the momentum fields $k^a$, ${\pi^i}_a$, and coordinate fields $q_a$, ${e_i}^a$, such that they satisfy Eqs.~\eqref{equg5.2} and \eqref{equg5.3}. The proposed generators $\tilde L_{0a}$ must satisfy the following conditions:

\begin{enumerate}[label={\arabic*)}]
\item 
$\tilde{L}_{0a}$ must consist of tensorial terms with a free spatial index "$a$", such as $k_a$, $q_a$, $q^e \tilde{M}_{ae}$, and similar terms.
\item
To satisfy Eq.~\eqref{equg5.2}, we require the Poisson bracket $\left\{\tilde{L}_{0a},\tilde{L}'_{0b}\right\}$ to contain the full antisymmetric combination $\tilde{L}_{ab} = \tilde{M}_{ab} + \tilde{N}_{ab}$. This suggests a structure such as $\tilde{L}_{0a} \sim \cdots k_a + \cdots q^e \tilde{M}_{ae}$. The alternative form involving $k^e \tilde{M}_{ae}$ is disfavored, as the weak equality \eqref{equc3.2} for $k^a$ would introduce considerable complexity.
\item 
To ensure the closure of the algebra as defined by Eqs.~\eqref{equg5.2} and \eqref{equg5.3}, higher-rank tensors such as $k^bq_bq^e\tilde{M}_{ae}$ must be avoided, as such expressions significantly complicate the algebra.
\item 
Scalar functions constructed from $q_a$ and $e_{ia}$ are allowed. However, the only scalar that can be formed from $e_{ia}$ alone is the constant value "$3$", whereas any scalar function of $q_aq^a$ can be regarded as a function of $\gamma$.
\end{enumerate}
Combining these requirements, we propose the following preliminary form of the boost generators
\begin{equation}
	\tilde{L}_{0a}=f(\gamma)k_a+g(\gamma)q^e\tilde{M}_{ae}.
	\label{equg5.4}
\end{equation}
We now determine the explicit forms of $f(\gamma)$ and $g(\gamma)$ by enforcing the algebraic relations \eqref{equg5.2} and \eqref{equg5.3}. First, we calculate the following bracket
\begin{align}
\left\{f(\gamma(q)),k'_d\right\}&=
\frac{\mathrm d f}{\mathrm d \gamma}
\frac{\partial \gamma}{\partial q^d}
\delta_{\mathbf{x}\mathbf{x'}}\notag\\
&={\gamma}^{-1}\frac{\mathrm d f}{\mathrm d \gamma} q_d
\delta_{\mathbf{x}\mathbf{x'}}.
\label{equg5.5}
\end{align}
Now let us proceed to the main expression $\left\{\tilde{L}_{0a},\tilde{L}'_{0b}\right\}$ as follows
\begin{align}
	\left\{\tilde{L}_{0a},\tilde{L}'_{0b}\right\}=
	\left\{fk_a,f'k'_b\right\}
	+\left\{fk_a,g'q'^e\tilde{M}'_{be}\right\}
	-\left\{f'k'_b,gq^e\tilde{M}_{ae}\right\}
	+\left\{gq^e\tilde{M}_{ae},g'q'^f\tilde{M}'_{bf}\right\}.
	\label{equg5.6}
\end{align}
After evaluating each term, we obtain
\begin{align}
  	\left\{\tilde{L}_{0a},\tilde{L}'_{0b}\right\}&=
  	-\gamma^{-1}f\frac{\mathrm d f}{\mathrm d \gamma}
  	\left(q_ak_b-q_bk_a\right)
  	\delta_{\mathbf{x}\mathbf{x'}}
  	+q^e\left(q_b\tilde{M}_{ae}-q_a\tilde{M}_{be}\right)
  	\left(\gamma^{-1}f\frac{\mathrm d g}{\mathrm d \gamma}
  	-g^2\right)
  	\delta_{\mathbf{x}\mathbf{x'}}
  	\notag\\&\quad
  	+\left(2fg+g^2\left(\gamma^2-1\right)\right)\tilde{M}_{ab}
  	\delta_{\mathbf{x}\mathbf{x'}}.
  	\label{equg5.7}
  \end{align}
  In order for Eq.~\eqref{equg5.7} to match Eq.~\eqref{equg5.2}, the following must hold 
  \begin{gather}
  	\gamma^{-1}f\frac{\mathrm d f}{\mathrm d \gamma}=1,
  	\label{equg5.8}\\
  	2fg+g^2\left(\gamma^2-1\right)=-1,
  	\label{equg5.9}\\
\gamma^{-1}f\frac{\mathrm d g}{\mathrm d \gamma}-g^2=0.
  	\label{equh5.10}
  \end{gather}
which yield $f=\gamma$ and $g=-(\gamma+1)^{-1}$. Fortunately, the last equation is consistent with the solution of the first two. Taking all these points into account, we introduce the following expression as a preliminary candidate for the boost generator
\begin{equation}
\tilde{L}_{0a}=
\gamma k_a-(\gamma+1)^{-1}q^e\tilde{M}_{ae}.
\label{equh5.11}
\end{equation}
So far, we have verified that $\tilde{L}_{0a}$ satisfies Eq.~\eqref{equg5.2}. We now verify that it also satisfies Eq.~\eqref{equg5.3}. Using Eqs.~\eqref{equh5.11} and \eqref{equf4.16}, we calculate
\begin{align}
\left\{\tilde{L}_{0a},\tilde{L}'_{cd}\right\}=
\left\{\gamma k_a,\tilde{N}'_{cd}\right\}
-(\gamma+1)^{-1}q^e
\left\{\tilde{M}_{ae},\tilde{M}'_{cd}\right\}
-\tilde{M}_{ae}\left\{(\gamma+1)^{-1}q^e
,\tilde{N}'_{cd}\right\}.
\label{equh5.12}
\end{align}
By explicit calculation, we find
\begin{gather}
\left\{\gamma k_a,\tilde{N}'_{cd}\right\}=
\left(-\eta_{ac}\gamma k_d+\eta_{ad}\gamma k_c\right)
\delta_{\mathbf{x}\mathbf{x'}},
\label{equh5.13}\\
-(\gamma+1)^{-1}q^e
\left\{\tilde{M}_{ae},\tilde{M}'_{cd}\right\}=
-(\gamma+1)^{-1}\left(\eta_{ac}q^e\tilde{M}_{ed}
-\eta_{ad}q^e\tilde{M}_{ec}+q_d\tilde{M}_{ac}
-q_c\tilde{M}_{ad}\right)
\delta_{\mathbf{x}\mathbf{x'}},
\label{equh5.14}\\
-\tilde{M}_{ae}\left\{(\gamma+1)^{-1}q^e
,\tilde{N}'_{cd}\right\}=
(\gamma+1)^{-1}\left(q_d\tilde{M}_{ac}
-q_c\tilde{M}_{ad}\right)
\delta_{\mathbf{x}\mathbf{x'}}.
\label{equh5.15}
\end{gather}
By substituting Eqs.~\eqref{equh5.13}--\eqref{equh5.15} into Eq.~\eqref{equh5.12}, we arrive at Eq.~\eqref{equg5.3}
\begin{align}
\left\{\tilde{L}_{0a},\tilde{L}'_{cd}\right\}&=
\left(\eta_{ad}\left(\gamma k_c
-(\gamma+1)^{-1}q^e\tilde{M}_{ce}\right)
-\eta_{ac}\left(\gamma k_d
-(\gamma+1)^{-1}q^e\tilde{M}_{de}\right)\right)
\delta_{\mathbf{x}\mathbf{x'}}
\notag\\&=
\left(\eta_{ad}\tilde{L}_{0c}-\eta_{ac}\tilde{L}_{0d}\right)
\delta_{\mathbf{x}\mathbf{x'}}.
\label{equh5.16}
\end{align}
\subsection{The main generators\texorpdfstring{ $\mathbf {L_{0a}}$}{PDFstring}}\label{L0a}
So far, we have constructed the preliminary boost generators $\tilde{L}_{0a}$. We now aim to obtain the main generators ${L}_{0a}$ by employing the weak equalities \eqref{equc3.2} and \eqref{equc3.5}. Using these, we define 
\begin{equation}
L_{0a}\equiv\tilde L_{0a}+2v_{0a}\approx0,
\label{equh5.17}
\end{equation}
where
\begin{equation}
v_{0a}=
\partial_i
\left((\gamma+1)^{-1}eq_aq^c{e^i}_c-\gamma e{e^i}_a\right).
\label{equh5.18}
\end{equation}
In this way, we propose that the weakly vanishing quantities $L_{0a}$ act as the main generators of the boost algebra. We shall examine this in the following.
\subsection{Boost algebra}\label{Boost algebra}
Let us now examine the boost-boost and boost-rotation sectors of the Lorentz algebra, as given in Eqs.~\eqref{equg5.2} and \eqref{equg5.3}. Using Eq.~\eqref{equh5.17}, we obtain
\begin{align}
\left\{L_{0a},L'_{0b}\right\}=
\left\{\tilde {L}_{0a},\tilde{L}'_{0b}\right\}
+2\left\{\tilde {L}_{0a},v'_{0b}\right\}
-2\left\{\tilde {L}'_{0b},v_{0a}\right\}
+4\left\{v_{0a},v'_{0b}\right\}.
\label{equh5.19}
\end{align}
The last term in Eq.~\eqref{equh5.19} vanishes since $v_{0a}$ depends only on the coordinate fields. Let us now evaluate the second term on the right-hand side of Eq.~\eqref{equh5.19}. We have
\begin{align}
\left\{\tilde {L}_{0a},v'_{0b}\right\}&=
\left\{\gamma k_a-(\gamma+1)^{-1}q^e\tilde{M}_{ae}
,\partial'_i\left((\gamma'+1)^{-1}e'q'_bq'^c{e'^i}_c
-\gamma' e'{e'^i}_b\right)\right\}
\notag\\&=
\partial'_i\left\{\gamma k_a
-(\gamma+1)^{-1}q^e\tilde{M}_{ae}
,\left((\gamma'+1)^{-1}e'q'_bq'^c{e'^i}_c
-\gamma' e'{e'^i}_b\right)\right\}.
\label{equh5.20}
\end{align}
Using the identity
$\left\{\tilde{M}_{a_1a_2}, {e'^i}_{a_3}\right\}=
\left(\eta_{a_1a_3}{e^i}_{a_2}-\eta_{a_2a_3}{e^i}_{a_1}\right)
\delta_{\mathbf{x}\mathbf{x'}}$,
Eq.~\eqref{equh5.20} simplifies to
\begin{align}
\left\{\tilde {L}_{0a},v'_{0b}\right\}=&
\partial'_i\left(e \left(q_a{e^i}_b-q_b{e^i}_a\right)
\delta_{\mathbf{x}\mathbf{x'}}\right)\notag\\
&=-e\left(q_a{e^i}_b-q_b{e^i}_a\right)
\partial_i\delta_{\mathbf{x}\mathbf{x'}}.
\label{equh5.21}
\end{align}
By exchanging the indices $a\leftrightarrow b$ and replacing $\partial'\rightarrow \partial$ in Eq.~\eqref{equh5.21}, we also obtain
\begin{align}
-\left\{\tilde {L}'_{0b},v_{0a}\right\}=
-\partial_i\left( e \left(q_b{e^i}_a-q_a{e^i}_b\right)
\delta_{\mathbf{x}\mathbf{x'}}\right).
\label{equh5.22}
\end{align}
Substituting Eqs.~\eqref{equh5.21}, \eqref{equh5.22}, \eqref{equf4.15}, and \eqref{equf4.17} into Eq.~\eqref{equh5.19}, we arrive at
\begin{align}
\left\{L_{0a},L'_{0b}\right\}&=
-\tilde{L}_{ab}\delta_{\mathbf{x}\mathbf{x'}}
-2\partial_i\left( e \left(q_b{e^i}_a-q_a{e^i}_b\right)
\right)\delta_{\mathbf{x}\mathbf{x'}}
\notag\\&=
-\left(\tilde{L}_{ab}+2v_{ab}\right)
\delta_{\mathbf{x}\mathbf{x'}}
\notag\\&=
\eta_{00}L_{ab}\delta_{\mathbf{x}\mathbf{x'}}.
\label{equh5.23}
\end{align}
\subsection{Boost-Rotation algebra}\label{Boost-Rotation algebra}
Let us calculate the Poisson bracket $\left\{L_{0a},L'_{cd}\right\}$ as follows
\begin{align}
\left\{L_{0a},L'_{cd}\right\}&=
\left\{\tilde{L}_{0a},\tilde{L}'_{cd}\right\}
+2\left\{\tilde{L}_{0a},v'_{cd}\right\}
-2\left\{\tilde{L}'_{cd},v_{0a}\right\}
+4\left\{v_{0a},v'_{cd}\right\}.
\label{equh5.24}
\end{align}
Again, the last term vanishes since both $v_{0a}$ and $v'_{cd}$ depend only on the coordinate fields and not on the momenta. To evaluate the second term on the right-hand side of Eq.~\eqref{equh5.24}, we write
\begin{align}
\left\{\tilde{L}_{0a},v'_{cd}\right\}&=
\left\{\gamma k_a-(\gamma+1)^{-1}q^e\tilde{M}_{ae}
,\partial'_i\left(e'\left(q'_d{e'^i}_c
-q'_c{e'^i}_d\right)\right)\right\}
\notag\\&=
-\left(\eta_{ad}\left((\gamma+1)^{-1}eq_cq^e{e^i}_e
-\gamma e{e^i}_c\right)
-\eta_{ac}\left((\gamma+1)^{-1}eq_dq^e{e^i}_e
-\gamma e{e^i}_d\right)\right)
\partial_i\delta_{\mathbf{x}\mathbf{x'}}.
\label{equh5.25}
\end{align}
For the third term in Eq.~\eqref{equh5.24}, we obtain
\begin{align}
-\left\{\tilde{L}'_{cd},v_{0a}\right\}&=
-\left\{\tilde{M}'_{cd}+\tilde{N}'_{cd}
,\partial_i\left((\gamma+1)^{-1}eq_aq^e{e^i}_e
-\gamma e{e^i}_a\right)\right\}
\notag\\&=
-\partial_i\left\{\tilde{M}'_{cd}+\tilde{N}'_{cd}
,\left((\gamma+1)^{-1}eq_aq^e{e^i}_e
-\gamma e{e^i}_a\right)\right\}.
\label{equh5.26}
\end{align}
By direct calculation, we find the following relations
\begin{align}
\left\{\tilde{M}'_{cd},
(\gamma+1)^{-1}eq_aq^e{e^i}_e-\gamma e{e^i}_a\right\}=&
(\gamma+1)^{-1}eq_a\left(q_c{e^i}_d-q_d{e^i}_c\right)
\delta_{\mathbf{x}\mathbf{x'}}\notag\\&
-\gamma e\left(\eta_{ac}{e^i}_d-\eta_{ad}{e^i}_c \right)
\delta_{\mathbf{x}\mathbf{x'}},
\label{equh5.27}
\end{align}
\begin{align}
\left\{\tilde{N}'_{cd}
,(\gamma+1)^{-1}eq_aq^e{e^i}_e-\gamma e{e^i}_a\right\}&=
(\gamma+1)^{-1}eq^e{e^i}_e
\left(-\eta_{ad}q_c+\eta_{ac}q_d\right)
\delta_{\mathbf{x}\mathbf{x'}}
\notag\\&\quad
+(\gamma+1)^{-1}eq_a\left(-q_c{e^i}_d
+q_d{e^i}_c\right)
\delta_{\mathbf{x}\mathbf{x'}}.
\label{equh5.28}
\end{align}
Since the term $\left\{\tilde{N}'_{cd},\gamma e{e^i}_a\right\}$ vanishes, Eq.~\eqref{equh5.28} arises solely from the term
$\left\{\tilde{N}'_{cd},(\gamma+1)^{-1}eq_aq^e{e^i}_e\right\}$.
Summing Eqs.~\eqref{equh5.27} and \eqref{equh5.28}, we obtain
\begin{align}
-\left\{\tilde{L}'_{cd},v_{0a}\right\}&=
\partial_i\left(\left(\eta_{ad}\left((\gamma+1)^{-1}
eq_cq^e{e^i}_e-\gamma e {e^i}_c\right)
-\eta_{ac}\left((\gamma+1)^{-1}eq_dq^e{e^i}_e
-\gamma e {e^i}_d\right)\right)
\delta_{\mathbf{x}\mathbf{x'}}\right).
\label{equh5.29}
\end{align}
Now, inserting the results from Eqs.~\eqref{equh5.16}, \eqref{equh5.25}, and \eqref{equh5.29} into Eq.~\eqref{equh5.24}, we find
\begin{align}
\left\{L_{0a},L'_{cd}\right\}&=
\eta_{ad}\left(\tilde{L}_{0c}
+2\partial_i\left((\gamma+1)^{-1}eq_cq^e{e^i}_e
-\gamma e {e^i}_c\right)\right)
\delta_{\mathbf{x}\mathbf{x'}}
\notag\\&\quad
-\eta_{ac}\left(\tilde{L}_{0d}
+2\partial_i\left((\gamma+1)^{-1}eq_dq^e{e^i}_e
-\gamma e {e^i}_d\right)\right)
\delta_{\mathbf{x}\mathbf{x'}}
\notag\\&=
\left(\eta_{ad}\left(\tilde{L}_{0c}+2v_{0c}\right)
-\eta_{ac}\left(\tilde{L}_{0d}+2v_{0d}\right)\right)
\delta_{\mathbf{x}\mathbf{x'}}
\notag\\&=
\left(\eta_{ad}L_{0c}-\eta_{ac}L_{0d}\right)
\delta_{\mathbf{x}\mathbf{x'}}.
\label{equh5.30}
\end{align}
Finally, we observe that Eqs.~\eqref{equf4.25}, \eqref{equh5.23}, and \eqref{equh5.30} confirm the existence of six weakly vanishing expressions $L_{AB}$, which serve as the generators of LLT in general relativity.
\section{Final remarks}\label{Final remarks}
Our final results for the generators of LLT are not simple enough to have been guessed at the outset. Indeed, we followed a long and challenging path to construct them. In the literature, the Hamiltonian (as well as Lagrangian) of general relativity is not written in terms of the complete set of 16 ADM-Vielbein variables. In fact, the triangular form of vielbein (Eqs.~\eqref{equa2.7} and \eqref{equa2.8}) is the maximum considered thing yet. It is often claimed that the boost variable $q_a$ (see Eqs.~\eqref{equa2.9} and \eqref{equb2.10}) plays no role in the Hamiltonian formalism. Even the generators of rotations considered in the literature are incomplete and contain only the simple components $\tilde{L}_{ab}$'s (see Eq.~\eqref{equf4.16}). As we observed, the necessity that the generators should be first class quantities, forces us to replace $\tilde{L}_{ab}$ with ${L}_{ab}\equiv\tilde{L}_{ab}+2v_{ab}$, where $v_{ab}$ are non-trivial combinations of coordinate fields (see Eq.~\eqref{equf4.17}).

The most creative part of this paper is the method of constructing the generators $L_{0a}$, which generate boost transformations. This concept has not been addressed in any previous references. In fact, it is not surprising that by ignoring the boost parameters $q_a$, one has already fixed the boost part of the Lorentz algebra, rendering it impossible to find the corresponding generators.

As mentioned in the introduction, by counting the number of degrees of freedom, it is clear that the primary constraints $L_{AB}$ can not lead to secondary constraints under the consistency conditions. However, a lengthy but straightforward calculation confirms that the constraints $L_{AB}$ have weakly vanishing Poisson brackets with the Hamiltonian.

Remember that for simplicity and focusing on LLT symmetry, we restrict our analysis to the case $N=1$ and $N^i=0$, i.e., mini-superspace.
 A natural extension of the current analysis is to consider the full set of dynamical variables, including the lapse and shift functions $N$ and $N^i$. By its very nature, the ADM formalism separates spacetime into space and time. This separation breaks explicit Lorentz covariance and makes it difficult to observe the full Poincaré invariance directly. For a more complete analysis, it is necessary to use the full (without gauge-fixing) Lagrangian and examine whether the resulting constraints truly reproduce the full Poincaré algebra. Such an analysis is significantly more complicated and lies beyond the scope of this paper. Nevertheless, we consider this line of investigation as a direction for future research.
\section*{\large{Acknowledgement}}
The authors are thankful for the cooperation of M. J. Vasli and G. Jafari for their collaboration in the initial stages of this research.
\begin{appendices}
\section{Some details}\label{Appendices A}
\subsection{The explicit form of the tensor 
\texorpdfstring{\boldmath$\Omega^{ABC}$}{PDFstring}}
\label{Appendix omegaABC}
Using the definition~\eqref{equb2.15} along with Eqs.~\eqref{equb2.19}, we explicitly obtain the following expressions for the components of $\Omega^{ABC}$
\begin{align}
\Omega^{0a0}=-\Omega^{a00}=&\left((\gamma+1)^{-1}{e^j}_cq^cq^a
-\gamma e^{ja}\right)q_d\partial_0{e_j}^d
+(\gamma+1)^{-1}q^aq^d\partial_0q_d-\gamma\partial_0q^a\notag\\
&-\gamma^{-1}(\gamma+1)^{-1}{e^j}_cq^cq^d\partial_jq_dq^a
+\left({e^j}_cq^ce^{ka}-{e^k}_cq^ce^{ja}\right)q_d\partial_j{e_k}^d
+{e^j}_cq^c\partial_jq^a,
\label{A.01}
\end{align}
\begin{align}
\Omega^{0ab}=-\Omega^{a0b}=&\left((\gamma+1)^{-1}{e^j}_cq^cq^a
-\gamma e^{ja}\right)\left(\partial_0{e_j}^b
+(\gamma+1)^{-1}q^bq_d\partial_0{e_j}^d\right)\notag\\
&+\gamma^{-1}(\gamma+1)^{-1}q^aq^bq^d\partial_0q_d
-(\gamma+1)^{-1}\left(\gamma q^b\partial_0q^a
+q^a\partial_0q^b\right)\notag\\
&+\left({e^j}_cq^ce^{ka}-{e^k}_cq^ce^{ja}\right)
\left(\partial_j{e_k}^b
+(\gamma+1)^{-1}q^bq_d\partial_j{e_k}^d\right)\notag\\
&+(\gamma+1)^{-1}{e^j}_cq^c\left(q^b\partial_jq^a
-\gamma^{-1}(\gamma+1)^{-1}q^aq^bq^d\partial_jq_d\right)\notag\\
&+e^{ja}\left(\partial_jq^b-\gamma^{-1}(\gamma+1)^{-1}
q^bq^d\partial_jq_d\right),
\label{A.02}
\end{align}
\begin{align}
\Omega^{ab0}=-\Omega^{ba0}=&\left(e^{ja}q^b-e^{jb}q^a\right)
q^c\partial_0{e_j}^c+q^b\partial_0q^a-q^a\partial_0q^b
+\left(e^{ja}e^{kb}-e^{jb}e^{ka}\right)q_c\partial_j{e_k}^c\notag\\
&+(\gamma+1)^{-1}\left({e^j}_dq^d\left(q^ae^{kb}-q^be^{ka}\right)
+{e^k}_dq^d\left(q^be^{ja}-q^ae^{jb}\right)\right)
q_c\partial_j{e_k}^c\notag\\
&+e^{ja}\partial_jq^b-e^{jb}\partial_jq^a
+(\gamma+1)^{-1}{e^j}_dq^d\left(q^a\partial_jq^b
-q^b\partial_jq^a\right)\notag\\
&-\gamma^{-1}(\gamma+1)^{-1}q^c\partial_jq_c
\left(e^{ja}q^b-e^{jb}q^a\right),
\label{A.03}
\end{align}
\begin{align}
\Omega^{abc}=-\Omega^{bac}=&\left(e^{ja}q^b-e^{jb}q^a\right)
\left(\partial_0{e_j}^c
+(\gamma+1)^{-1}q^cq_d\partial_0{e_j}^d\right)
+(\gamma+1)^{-1}q^c\left(q^b\partial_0q^a-q^a\partial_0q^b\right)
\notag\\
&+(\gamma+1)^{-1}\left({e^j}_dq^d\left(q^ae^{kb}-q^be^{ka}\right)
+{e^k}_dq^d\left(q^be^{ja}-q^ae^{jb}\right)\right)
\left(\partial_j{e_k}^c
+(\gamma+1)^{-1}q^cq_e\partial_j{e_k}^e\right)\notag\\
&+\left(e^{ja}e^{kb}-e^{jb}e^{ka}\right)
\left(\partial_j{e_k}^c
+(\gamma+1)^{-1}q^cq_d\partial_j{e_k}^d\right)
-(\gamma+1)^{-1}\partial_jq^c\left(q^be^{ja}-q^ae^{jb}\right)\notag\\
&+(\gamma+1)^{-1}q^c\left(e^{ja}\partial_jq^b
-e^{jb}\partial_jq^a\right)
+(\gamma+1)^{-2}{e^j}_dq^dq^c\left(q^a\partial_jq^b
-q^b\partial_jq^a\right).
\label{A.04}
\end{align}
\subsection{The explicit form of \texorpdfstring{\boldmath$\mathcal{L}_{EH}$}{PDFstring}}
\label{Appendix l_EH}
In this appendix, we determine $\mathcal{L}_{EH}$ in the mini-superspace setting $(N=1$ and $N^i=0)$. In addition to Eqs.~\eqref{equb2.18}, \eqref{equb2.19}, \eqref{A.01}, \eqref{A.02}, \eqref{A.03}, and \eqref{A.04}, the following identities are also used in determining $\mathcal{L}_{EH}$
\begin{gather}
	{E_\mu}^A\partial_\rho{E^\nu}_A=
	-{E^\nu}_A\partial_\rho{E_\mu}^A\,,\qquad
	{E_\mu}^A\partial_\rho{E^\mu}_B=
	-{E^\mu}_B\partial_\rho{E_\mu}^A,\label{A.05}\\
	E_{\mu A}\partial_\rho{E^\mu}_B=
	-{E^\mu}_B\partial_\rho E_{\mu A}\,,\qquad 
	{E_\mu}^A\partial_\rho E^{\mu B}=
	-E^{\mu B}\partial_\rho {E_\mu}^A,\label{A.06} \\
	{e_l}^b\partial_\rho {e^k}_b=
	-{e^k}_b\partial_\rho{e_l}^b\,,\qquad
	{e_k}^d\partial_\rho{e^k}_b=
	-{e^k}_b\partial_\rho{e_k}^d,\label{A.07}\\
	e_{i a}\partial_\rho{e^i}_b=
	-{e^i}_b\partial_\rho e_{i a}\,,\qquad 
	{e_i}^a\partial_\rho e^{i b}=
	-e^{i b}\partial_\rho {e_i}^a.\label{A.08}
\end{gather}
We also frequently use the identity $q^cq_c=\gamma^2-1$. We now express the Lagrangian density \eqref{equb2.17} as
\begin{equation}
	\mathcal{L}_{EH}=\mathcal{L}_{0}+\mathcal{L}_{1}+\mathcal{L}_{2}.
	\label{A.09}
\end{equation}
Here, $\mathcal{L}_2$ is quadratic, $\mathcal{L}_1$ is linear in the velocities, and $\mathcal{L}_0$ contains no temporal derivatives. After carrying out detailed calculations, we arrive at the following results
\begin{align}
	\mathcal{L}_2=
	e\left({e^l}_c\partial_0{e_l}^c{e^m}_b\partial_0{e_m}^b
	+\frac{1}{2}\partial_0{e^l}_b\partial_0{e_l}^b
	-\frac{1}{2}e^{mb}{e^l}_b
	\partial_0{e_l}^c\partial_0 e_{mc}\right),
	\label{mathcal{L}_2}
\end{align}
\begin{align}
	\mathcal{L}_1&=2e\left(
	-\gamma^{-1}(\gamma+1)^{-1}q^c\partial_0q_c{e^j}_d\partial_jq^d
	+\gamma^{-1}(\gamma+1)^{-1}q^c\partial_jq_c{e^j}_d\partial_0q^d
	\right.
	\notag\\&\hspace{1.15 cm}
	-\gamma^{-1}(\gamma+1)^{-1}q^c\partial_0q_cq^d\partial_j{e^j}_d	+\gamma^{-1}(\gamma+1)^{-1}q^c\partial_jq_cq^d\partial_0{e^j}_d
	\notag\\&\hspace{1.15 cm}
	+\partial_0q^c\partial_j{e^j}_c
	-\partial_jq^c\partial_0{e^j}_c
	-\gamma^{-1}(\gamma+1)^{-1}{e^j}_bq^bq^d\partial_0q_d
	{e^l}_c\partial_j{e_l}^c
	\notag\\&\hspace{1.15 cm}
	\left.
	+{e^j}_d\partial_0q^d{e^l}_c\partial_j{e_l}^c
	+\gamma^{-1}(\gamma+1)^{-1}{e^j}_bq^bq^d\partial_jq_d
	{e^l}_c\partial_0{e_l}^c
	-{e^j}_d\partial_jq^d{e^l}_c\partial_0{e_l}^c
	\right),
	\label{mathcal{L}_1}
\end{align}
\begin{align}
	\mathcal{L}_0&=e\left(-{e^j}_be^{kb}
	{e^l}_c\partial_j{e_l}^c{e^m}_d\partial_k{e_m}^d
	-\frac{1}{2}{e^j}_be^{kb}
	\partial_j{e_l}^c\partial_k{e^l}_c
	+\frac{1}{2}{e^j}_be^{kb}e^{md}{e^l}_d
	\partial_j{e_l}^c\partial_ke_{mc}\right.
	\notag\\&\hspace{1 cm}
	+2(\gamma+1)^{-1}{e^k}_bq^b
	{e^j}_d\partial_jq^d{e^l}_c\partial_k{e_l}^c
	-2e^{kb}\partial_j{e^j}_b{e^l}_c\partial_k{e_l}^c
	\notag\\&\hspace{1 cm}
	-2(\gamma+1)^{-1}{e^j}_bq^b{e^k}_d\partial_jq^d
	{e^l}_c\partial_k{e_l}^c
	-\frac{1}{2}{e^j}_be^{mb}{e^k}_ce^{lc}
	\partial_j{e_l}^d\partial_ke_{md}
	\notag\\&\hspace{1 cm}
	+{e^k}_ce^{lc}\partial_j{e_l}^b
	\partial_k{e^j}_b
	+\frac{1}{2}\partial_j{e^k}_c\partial_ke^{jc}
	-\partial_k{e^k}_c\partial_je^{jc}
	\notag\\&\hspace{1 cm}
	-2(\gamma+1)^{-1}{e^k}_c\partial_jq^c
	{e^j}_d\partial_kq^d
	+2(\gamma+1)^{-1}{e^k}_c\partial_kq^c
	{e^j}_d\partial_jq^d
	\notag\\&\hspace{1 cm}
	-2(\gamma+1)^{-1}{e^k}_c\partial_jq^c
	q^d\partial_k{e^j}_d
	+2(\gamma+1)^{-1}{e^k}_c\partial_kq^c
	q^d\partial_j{e^j}_d
	\notag\\&\hspace{1 cm}
	+2(\gamma+1)^{-1}{e^j}_bq^b
	\partial_j{e^k}_c\partial_kq^c
	-2(\gamma+1)^{-1}{e^j}_bq^b
	\partial_k{e^k}_c\partial_jq^c
	\notag\\&\hspace{1 cm}
	\left.
	+2\gamma^{-1}(\gamma+1)^{-2}{e^k}_dq^d
	{e^j}_b\partial_kq^bq^c\partial_jq_c
	-2\gamma^{-1}(\gamma+1)^{-2}{e^k}_dq^d
	{e^j}_b\partial_jq^bq^c\partial_kq_c
	\vphantom{\frac{1}{2}}\right).
	\label{mathcal{L}_0}
\end{align}
\subsection{Determination of the momentum \texorpdfstring{\boldmath{${\pi^i}_a$}}{PDFstring}}\label{Appendix pi^i_a}
The momentum ${\pi^i}_a$ can be derived from the explicit form of $\mathcal{L}_{EH}$ via
\begin{align}
	{\pi^i}_a&=\frac{\partial\mathcal{L}_{EH}}
	{\partial(\partial_0{e_i}^a)}\notag\\
	&=\frac{\partial\mathcal{L}_{2}}
	{\partial(\partial_0{e_i}^a)}+\frac{\partial\mathcal{L}_{1}}
	{\partial(\partial_0{e_i}^a)}.
	\label{A.13}
\end{align}
The following equations are used in determining ${\pi^i}_a$
\begin{gather}
	\frac{\partial(\partial_0{e_l}^b)}
	{\partial(\partial_0{e_i}^a)}=\delta^i_l\delta^b_a,\label{A.14}\\
	\frac{\partial(\partial_0{e^l}_b)}
	{\partial(\partial_0{e_i}^a)}=-{e^i}_b{e^l}_a.
	\label{A.15}
\end{gather}
Eq.~\eqref{A.15} is not immediately obvious. Contracting both sides of Eq.~\eqref{A.07} with ${e^l}_d$ gives 
\begin{equation}
	\partial_\mu {e^l}_b=-{e^l}_d{e^k}_b\partial_\mu{e_k}^d.
	\label{A.16}
\end{equation}
Eq.~\eqref{A.15} follows from Eqs.~\eqref{A.14} and \eqref{A.16}.
Using equations \eqref{mathcal{L}_2}, \eqref{mathcal{L}_1}, \eqref{A.13}, \eqref{A.14}, and \eqref{A.15}, one can determine ${\pi^i}_a$ as given in Eq.~\eqref{equc3.1}.

Alternatively, the momentum ${\pi^i}_a$ can be determined without using the explicit form of the Lagrangian density as follows
\begin{align}
	{\pi^i}_a&=
	\frac{\partial\mathcal{L}_{EH}}
	{\partial(\partial_0{e_i}^a)}
	\notag\\
	&=e\left(\frac{1}{2}\Omega_{ABC}
	\frac{\partial\Omega^{ABC}}{\partial(\partial_0{e_i}^a)}
	+\Omega_{ACB}
	\frac{\partial\Omega^{ABC}}{\partial(\partial_0{e_i}^a)}
	-2{\Omega_B}^{CB}\frac{\partial{\Omega_{AC}}^A}
	{\partial(\partial_0{e_i}^a)}\right).
	\label{A.17}
\end{align}
In addition to the previously mentioned relations, the following equations are also employed in determining the various terms of Eq.~\eqref{A.17}
\begin{gather}
	\frac{\partial(\partial_0 {E_j}^0)}
	{\partial(\partial_0{e_i}^a)}=q_a\delta^i_j\label{A.18},\\
	\frac{\partial(\partial_0 {E_j}^b)}
	{\partial(\partial_0{e_i}^a)}=\delta^i_j\delta^b_a
	+(\gamma+1)^{-1}q_aq^b\delta^i_j\label{A.19}.
\end{gather}
For various terms of the Eq.~\eqref{A.17} we obtain the following expressions
\begin{align}
	\frac{1}{2}\Omega_{ABC}\frac{\partial\Omega^{ABC}}{\partial
		(\partial_0{e_i}^a)}
	=&\left(g^{\rho \nu}\partial_\rho E_{0C}
	-g^{\nu \sigma}\partial_0 E_{\sigma C}\right)
	\frac{\partial(\partial_0 {E_\nu}^C)}
	{\partial(\partial_0{e_i}^a)}\notag\\
	=&-\gamma^{-1}(\gamma+1)^{-1}q_ae^{ib}{e^j}_bq_c\partial_jq^c
	-(\gamma+1)^{-1}q^b{e^i}_b\partial_0q_a
	\notag\\
	&+(\gamma+1)^{-1}q_a{e^i}_b\partial_0q^b
	+e^{ib}{e^j}_b\partial_jq_a-e^{ib}{e^j}_b\partial_0e_{ja},
	\label{A.20}
\end{align}
\begin{align}
	\Omega_{ACB}\frac{\partial\Omega^{ABC}}{\partial
		(\partial_0{e_i}^a)}=&
	\left(-g^{0 \rho}\partial_\rho{E^\nu}_C
	+{E^\rho}_C{E^0}_B\partial_\rho E^{\nu B}
	+g^{\nu \rho}\partial_\rho {E^0}_C
	-{E^\rho}_C{E^\nu}_B\partial_\rho E^{0 B}\right)
	\frac{\partial(\partial_0 {E_\nu}^C)}
	{\partial(\partial_0{e_i}^a)}\notag\\
	=&\gamma^{-1}(\gamma+1)^{-1}q_a{e^i}_ce^{jc}q_b\partial_j q^b
	-(\gamma+1)^{-1}q_a{e^i}_b\partial_0q^b
	+\partial_0{e^i}_a
	-{e^i}_ce^{jc}\partial_jq_a
	\notag\\
	&+(\gamma+1)^{-1}q^c{e^i}_c\partial_0q_a
	+2{e^j}_a{e^i}_c\partial_jq^c	
	-2\gamma^{-1}(\gamma+1)^{-1}{e^j}_aq^c{e^i}_cq_b\partial_jq^b,
	\label{A.21}
\end{align}
\begin{align}
	-2{\Omega_B}^{CB}
	\frac{\partial{\Omega_{AC}}^A}{\partial(\partial_0{e_i}^a)}&=
	2\left(\left({E^0}_AE^{\nu B}
	-{E^\nu}_AE^{0 B}\right)\partial_\rho{E^\rho}_B
	+\left(g^{\rho \nu}{E^0}_A
	-g^{\rho 0}{E^\nu}_A\right){E^\sigma}_B
	\partial_\rho {E_\sigma}^B\right)
	\frac{\partial(\partial_0 {E_\nu}^A)}
	{\partial(\partial_0{e_i}^a)}\notag\\
	&=-2{e^i}_a{e^k}_b\partial_kq^b
	+2\gamma^{-1}(\gamma+1)^{-1}{e^i}_aq^d{e^k}_dq_b\partial_kq^b
	+2{e^i}_a{e^k}_b\partial_0 {e_k}^b.
	\label{A.22}
\end{align}
Eqs.~\eqref{A.17}, \eqref{A.20}, \eqref{A.21}, and \eqref{A.22} lead to the same expression for ${\pi^i}_a$ as given in Eq.~\eqref{equc3.1}.
\subsection{Determination of the momentum \texorpdfstring{\boldmath{$k^a$}}{PDFstring}}\label{Appendix k^a}
Similar to ${\pi^i}_a$, the momentum $k^a$ can also be determined without using the explicit form of the Lagrangian density. For this purpose, we have
\begin{align}
	k^a&=
	e\left(\frac{1}{2}\Omega_{ABC}
	\frac{\partial\Omega^{ABC}}{\partial(\partial_0q_a)}
	+\Omega_{ACB}\frac{\partial\Omega^{ABC}}
	{\partial(\partial_0q_a)}
	-2{\Omega_B}^{CB}
	\frac{\partial{\Omega_{AC}}^A}{\partial(\partial_0q_a)}\right).
	\label{A.23}
\end{align}
In addition to the previously mentioned relations, the following expressions are also used to evaluate various terms in Eq.~\eqref{A.23}
\begin{gather}
	\frac{\partial(\partial_0{E_j}^0)}
	{\partial(\partial_0q_a)}={e_j}^a,\label{A.24}\\
	\frac{\partial(\partial_0 {E_j}^d)}
	{\partial(\partial_0q_a)}=
	-\gamma^{-1}(\gamma+1)^{-2}q^a{e_j}^bq_bq^d
	+(\gamma+1)^{-1}q^d{e_j}^a
	+(\gamma+1)^{-1}\eta^{ad}{e_j}^bq_b.\label{A.25}
\end{gather}
The following results are obtained directly
\begin{align}
	\frac{1}{2}\Omega_{ABC}
	\frac{\partial\Omega^{ABC}}{\partial(\partial_0q_a)}
	=&\left(g^{\rho \nu}\partial_\rho E_{0C}
	-g^{\nu \sigma}\partial_0 E_{\sigma C}\right)
	\frac{\partial(\partial_0 {E_\nu}^C)}
	{\partial(\partial_0q_a)}\notag\\
	=&-\gamma^{-1}(\gamma+1)^{-1}e^{ja}q_b\partial_jq^b
	+(\gamma+1)^{-1}e^{ja}q^b\partial_0{e_j}^b
	+(3-\gamma)(\gamma+1)^{-1}\partial_0q^a
	\notag\\
	&-\gamma^{-1}(\gamma+1)^{-2}q^aq^b{e^j}_bq_c\partial_jq^c
	+(\gamma+1)^{-1}q^b{e^j}_b\partial_jq^a
	-(\gamma+1)^{-1}q^b{e^j}_b\partial_0{e_j}^a
	\notag\\
	&+(\gamma^2-2\gamma-1)
	\gamma^{-2}(\gamma+1)^{-2}q^aq_b\partial_0q^b,
	\label{A.26}
\end{align}
\begin{align}
	\Omega_{ACB}
	\frac{\partial\Omega^{ABC}}{\partial(\partial_0q_a)}=&
	\left(-g^{0 \rho}\partial_\rho{E^\nu}_C
	+{E^\rho}_C{E^0}_B\partial_\rho E^{\nu B}
	+g^{\nu \rho}\partial_\rho {E^0}_C
	-{E^\rho}_C{E^\nu}_B\partial_\rho E^{0 B}\right)
	\frac{\partial(\partial_0 {E_\nu}^C)}
	{\partial(\partial_0q_a)}\notag\\=&
	3\gamma^{-1}(\gamma+1)^{-1}e^{ja}q_b\partial_jq^b
	-(\gamma+1)^{-1}{e_j}^aq^b\partial_0{e^j}_b
	+(3\gamma-1)(\gamma+1)^{-1}\partial_0q^a
	\notag\\&
	-(3\gamma^2+2\gamma+1)
	\gamma^{-2}(\gamma+1)^{-2}q^aq_b\partial_0q^b
	-3(\gamma+1)^{-1}q^b{e^j}_b\partial_jq^a
	\notag\\&
	+3\gamma^{-1}(\gamma+1)^{-2}q^aq^b{e^j}_bq_c\partial_jq^c
	+(\gamma+1)^{-1}q_b{e_j}^b\partial_0e^{ja},
	\label{A.27}
\end{align}
\begin{align}
	-2{\Omega_B}^{CB}
	\frac{\partial{\Omega_{AC}}^A}{\partial(\partial_0q_a)}=&
	2\left(\left({E^0}_AE^{\nu B}
	-{E^\nu}_AE^{0 B}\right)\partial_\rho{E^\rho}_B
	+\left(g^{\rho \nu}{E^0}_A
	-g^{\rho 0}{E^\nu}_A\right){E^\sigma}_B
	\partial_\rho {E_\sigma}^B\right)
	\frac{\partial(\partial_0 {E_\nu}^A)}
	{\partial(\partial_0{e_i}^a)}\notag\\=&
	-2\partial_0q^a+2\gamma^{-2}q^aq_b\partial_0q^b
	+2\partial_je^{ja}
	+2(\gamma+1)^{-1}q^b{e^j}_b\partial_jq^a
	\notag\\&
	-2\gamma^{-1}(\gamma+1)^{-2}q^aq^b{e^j}_bq_c\partial_jq^c
	-2\gamma^{-1}(\gamma+1)^{-1}q^a{e^j}_b\partial_jq^b
	\notag\\&
	+2e^{ja}{e^k}_b\partial_j{e_k}^b
	-2\gamma^{-1}(\gamma+1)^{-1}
	q^aq^b\partial_j{e^j}_b
	\notag\\&
	-2\gamma^{-1}(\gamma+1)^{-1}
	q^aq^b{e^j}_b{e^k}_c\partial_j{e_k}^c.
	\label{A.28}
\end{align}
Hence, the momentum $k^a$ can be obtained by substituting Eqs.~\eqref{A.26}, \eqref{A.27}, and \eqref{A.28} into Eq.~\eqref{A.23}, as given in Eq.~\eqref{equc3.2}.

The same result for $k^a$ can also be derived from the explicit form of the Lagrangian density by noting that only $\mathcal{L}_1$ contains $\partial_0 q_a$. A direct calculation then leads to the same result as Eq.~\eqref{equc3.2}.
\subsection{Calculating two Poisson brackets}\label{Appendix two pb}
Except for the Poisson brackets given in Eqs.~\eqref{equd3.19} and \eqref{equd3.20}, the rest of the Poisson brackets in Eqs.~\eqref{equd3.16} to \eqref{equd3.27} are straightforward to calculate.

To derive Eq.~\eqref{equd3.19}, we use the identity ${e^i}_c {e_l}^c = \delta^i_l$ and proceed as follows
	\begin{gather*}
		\left\{{e^i}_c{e_l}^c,{\pi'^j}_b\right\}=0,\\
		{e^i}_c\delta_l^j\delta_b^c\delta_{\mathbf{x}\mathbf{x'}}
		+{e_l}^c\left\{{e^i}_c,{\pi'^j}_b\right\}=0,\\
		{e^l}_a{e^i}_c\delta_l^j\delta_b^c\delta_{\mathbf{x}\mathbf{x'}}
		+{e^l}_a{e_l}^c\left\{{e^i}_c,{\pi'^j}_b\right\}=0,\\
		\left\{{e^i}_a,{\pi'^j}_b\right\}
		=-{e^j}_a{e^i}_b\delta_{\mathbf{x}\mathbf{x'}}.\\
	\end{gather*}
To prove Eq.~\eqref{equd3.20}, we use the definition 
$e = {e_1}^b {e_2}^c {e_3}^d \epsilon_{bcd}$ and calculate
	\begin{align*}
		\left\{e,{\pi'^i}_a\right\}&=
		\left\{{e_1}^b{e_2}^c{e_3}^d \epsilon_{bcd},{\pi'^i}_a\right\}\\
		&=\left(\delta^i_1{e_2}^c{e_3}^d \epsilon_{acd}+
		\delta^i_2{e_1}^b{e_3}^d \epsilon_{bad}
		+\delta^i_3{e_1}^b{e_2}^c \epsilon_{bca}\right) \delta_{\mathbf{x}\mathbf{x'}}\\
		&=\left({e^i}_b{e_1}^b{e_2}^c{e_3}^d \epsilon_{acd}+
		{e^i}_c{e_2}^c{e_1}^b{e_3}^d \epsilon_{bad}
		+{e^i}_d{e_3}^d{e_1}^b{e_2}^c \epsilon_{bca}\right) \delta_{\mathbf{x}\mathbf{x'}}\\
		&={e_1}^b{e_2}^c{e_3}^d
		\left({e^i}_b \epsilon_{acd}+
		{e^i}_c\epsilon_{bad}
		+{e^i}_d\epsilon_{bca}\right)
		\delta_{\mathbf{x}\mathbf{x'}}\\
		&={e_1}^b{e_2}^c{e_3}^d{e^i}_a\epsilon_{bcd}
		\delta_{\mathbf{x}\mathbf{x'}}\\
		&=e{e^i}_a\delta_{\mathbf{x}\mathbf{x'}}.
	\end{align*}
\end{appendices}
\printbibliography
\end{document}